\documentclass[%
reprint,
superscriptaddress,
showpacs,
amsmath,amssymb,
pra,
]{revtex4-1}
\usepackage{graphicx}
\usepackage{dcolumn}
\usepackage{bm}
\usepackage{CJKutf8}
\usepackage{color}
\usepackage{amssymb}
\usepackage{amsfonts}
\usepackage[colorlinks,urlcolor=blue,linkcolor=blue,citecolor=blue,anchorcolor=blue]{hyperref}
\makeatletter

\newcommand{\Rmnum}[1]{\expandafter\@slowromancap\romannumeral #1@}
\makeatother
\begin{document}
	
	\begin{CJK*}{UTF8}{gbsn}
		
		\title{Floquet edge solitons in modulated trimer waveguide arrays}
		
		\author{Shuang Shen (沈双)}
		\affiliation{Key Laboratory for Physical Electronics and Devices of the Ministry of Education \& Shaanxi Key Lab of Information
			Photonic Technique, School of Electronic and Information Engineering, Xi'an Jiaotong University, Xi'an 710049, China}%
		
		\author{Yaroslav V. Kartashov}
		\affiliation{Institute of Spectroscopy, Russian Academy of Sciences, Troitsk, Moscow, 108840, Russia}
		
		\author{Yongdong Li (李永东)}
		\affiliation{Key Laboratory for Physical Electronics and Devices of the Ministry of Education \& Shaanxi Key Lab of Information
			Photonic Technique, School of Electronic and Information Engineering, Xi'an Jiaotong University, Xi'an 710049, China}%
		
		\author{Yiqi Zhang (张贻齐)}
		\email{zhangyiqi@xjtu.edu.cn}
		\affiliation{Key Laboratory for Physical Electronics and Devices of the Ministry of Education \& Shaanxi Key Lab of Information
			Photonic Technique, School of Electronic and Information Engineering, Xi'an Jiaotong University, Xi'an 710049, China}%
		
		\date{\today}
		
		\begin{abstract}
			\noindent
			We show that one-dimensional Floquet trimer arrays with periodically oscillating waveguides support two different and co-existing types of topological Floquet edge states in two different topological gaps in Floquet spectrum. In these systems nontrivial topology is introduced by longitudinal periodic oscillations of the waveguide centers, leading to the formation of Floquet edge states in certain range of oscillation amplitudes despite the fact that the structure spends half of the period in ``instantaneously'' nontopological phase, and only during other half-period it is ``instantaneously'' topological. Two co-existing Floquet edge states are characterized by different phase relations between bright spots in the unit cell -- in one mode these spots are in-phase, while in other mode they are out-of-phase. We show that in focusing nonlinear medium topological Floquet edge solitons, representing exactly periodic nonlinear localized Floquet states, can bifurcate from both these types of linear edge states. Both types of Floquet edge solitons can be stable and can be created dynamically using two-site excitations.
		\end{abstract}
		
		\maketitle
		
	\end{CJK*}
	
	\section{Introduction}
	
	The Floquet engineering is a mechanism that employs periodic modulations of parameters of the physical system in time (evolution variable)~\cite{rudner.prx.3.031005.2013, rudner.nrp.2.229.2020, kitagawa.prb.82.235114.2010} to generate new effects or novel phases of matter that are not observed in analogous static (unmodulated) systems. Such modulations lead to unusual interactions and coupling of eigenstates of the system, qualitative change of dispersion characteristics, unusual propagation paths for excitations, or even appearance of new types of the localized states, to name just a few applications. Floquet engineering is frequently used for manipulation of classical waves~\cite{yin.elight.2.8.2022}, in particular, in photonic systems, such as shallow waveguide arrays~\cite{szameit.jpb.43.163001.2010, garanovich.pr.518.1.2012, kartashov.nrp.1.185.2019}, where periodic variations of the optical properties in the direction of light propagation can be easily achieved, and where mathematical description of paraxial propagation of light waves is equivalent to the description of evolution of quantum particles in various time-modulated potentials.
	
	Periodic longitudinal modulations in Floquet waveguiding systems may create artificial gauge fields that break time-reversal symmetry in governing evolution equations and enable observation of a rich variety of phenomena of topological origin, such as photonic Floquet topological insulators~\cite{rechtsman.nature.496.196.2013}, anomalous insulators supporting unidirectional topological edge states for sufficiently large periods of modulation~\cite{maczewsky.nc.8.13756.2017, mukherjee.nc.8.13918.2017}, creation of bimorphic topological insulators~\cite{pyrialakos.nm.21.634.2022}, 
	various resonant coupling and tunneling mechanisms for topological edge states~\cite{zhang.lpr.12.1700348.2018, zhang.apl.3.120801.2018, zhong.ol.44.3342.2019, liu.prr.4.L012043.2022}, they enable new mechanisms of localization for light~\cite{lumer.np.13.339.2019},  and they may result in qualitative transformation of spectra of Floquet systems~\cite{leykam.prl.117.013902.2016}, to name just a few. When nonlinearity is introduced in topological Floquet waveguiding systems, a rich nomenclature of Floquet solitons in the topological bandgap can emerge~\cite{lumer.prl.111.243905.2013, leykam.prl.117.143901.2016, ablowitz.pra.90.023813.2014, ablowitz.pra.99.033821.2019, ivanov.acs.7.735.2020, ivanov.ol.45.1459.2020, ivanov.ol.45.2271.2020, mukherjee.prx.11.041057.2021, hang.pra.103.L040202.2021}, and nonlinearity-induced topological phases can be observed~\cite{maczewsky.science.370.701.2020}. 
	It is worth mentioning that among recently discovered topological effects in Floquet waveguiding systems is the formation of so-called anomalous $\pi$ edge modes~\cite{zhang.acs.4.2250.2017, cheng.prl.122.173901.2019, petracek.pra.101.033805.2020, wu.prr.3.023211.2021, song.lpr.15.2000584.2021, sidorenko.prr.4.033184.2022, zhong.pra.107.L021502.2023} that were first discovered in condensed matter physics~\cite{asboth.prb.90.125143.2014, dallago.pra.92.023624.2015, fruchart.prb.93.115429.2016}. These Floquet edge states arise in the simplest Su-Schrieffer-Heeger (SSH) arrays consisting of dimers (pairs of waveguides) when inter-cell and intra-cell couplings are periodically modulated in the direction of light propagation, so that the array switches between ``instantaneously'' topological and nontopological phases. 
	Beyond photonic platforms, $\pi$ modes were also observed in acoustic systems~\cite{cheng.prl.129.254301.2022, zhu.nc.13.11.2022}.
	A survey of recent results on topological effects in various Floquet systems can be found in reviews~\cite{lu.np.8.821.2014, ozawa.rmp.91.015006.2019, smirnova.apr.7.021306.2020, segev.nano.10.425.2021}.
	
	We would like to note that all one-dimensional photonic Floquet systems considered so far were based on the usual SSH arrays consisting of dimers~\cite{zhu.pra.98.013855.2018}, where only one topological bandgap opens in the Floquet spectrum. At the same time, extended static SSH-like systems have been found to be characterized by much richer spectra than their counterparts consisting of dimers~\cite{he.jpcm.33.085501.2021}. 
	Such systems include static trimer arrays~\cite{jin.pra.96.032103.2017, midya.pra.98.043838.2018, martinez.pra.99.013833.2019, zhang.oe.29.42827.2021, guo.njp.24.063001.2022, anastasiadis.prb.106.085109.2022}, where formation of nonlinear topological modes was reported very recently~\cite{kartashov.prl.128.093901.2022}. Extended topological Floquet SSH-like lattices were not addressed neither in linear nor in nonlinear regime, to the best of our knowledge.
	
	The goal of this work is to introduce linear Floquet edge states and bifurcating from them Floquet edge solitons in longitudinally modulated extended SSH arrays. To achieve this goal we consider the array consisting of trimers of waveguides, whose centers periodically oscillate in the direction of light propagation, mimicking time-modulations of lattice in condensed-matter Floquet systems. We show that the most representative feature of this photonic structure is the co-existence of two different types (in-phase one and out-of-phase one) of topological Floquet edge states with different internal structure in two topological gaps opening in the Floquet spectrum of the system due to longitudinal variations of its parameters. We also show that in the presence of focusing nonlinearity Floquet edge solitons may bifurcate from linear Floquet edge states in both topological gaps. The locations of quasi-propagation constants of such solitons in the gap and their stability depend on their power, but remarkable fact is that both types of topological solitons can be stable.
	
	While nonlinear Floquet topological systems have been considered in some previous publications, very few of them addressed the systems, where waveguides in the unit cell show different periodic evolution with distance, allowing to simultaneously open several topological gaps in the spectrum. Moreover, in clear contrast to previous works on $\pi$ modes utilizing dimers, in trimer arrays considered here such states appear not at the boundaries of the longitudinal Brillouin zone, but in its depth. Our results clarify the physics of Floquet topological systems based on lattices with complex internal structure, beyond usual dimerized lattices, and may have potential applications in development of Floquet topological lasers~\cite{ivanov.apl.4.126101.2019}, including lasers based on higher-order Floquet insulators, for enhancement of nonlinear interactions and harmonic generation processes involving topological edge states, photon entanglement~\cite{rechtsman.optica.3.925.2016}, and construction of on-chip compact topological optical devices~\cite{yin.elight.2.8.2022, chen.elight.1.2.2021}.
	
	\section{Linear Floquet edge states}\label{linear}
	
	Propagation of light beams in the extended oscillating SSH waveguide array inscribed in a focusing nonlinear medium can be described by the nonlinear Schr\"odinger-like equation:
	\begin{equation}\label{eq1}
		i \frac{\partial \psi}{\partial z} = -\frac{1}{2} \left( \frac{\partial^2}{\partial x^2} + \frac{\partial^2}{\partial y^2} \right) \psi
		-\mathcal{R}(x,y,z) \psi-|\psi|^{2} \psi,
	\end{equation}
	where $\psi$ is the dimensionless complex amplitude of the field, $x$ and $y$ are the normalized transverse coordinates, $z$ is the normalized propagation distance, and the function $\mathcal{R}(x,y,z)=\mathcal{R}(x,y,z+Z)$ describes $z$-periodic waveguide array with longitudinal period $Z$. The array consists of trimers, where centers of two outermost waveguides in each trimer periodically oscillate with distance $z$ 
	\begin{equation}\label{eq2}
		\mathcal{R}(x,y,z) = p \sum_m 
		\left(
		\begin{split}
			&e^{ -[(x-x_{m-1})^2+y^2]/\sigma^2} +\\
			&e^{ -[(x-x_m)^2+y^2]/\sigma^2} +\\
			&e^{ -[(x-x_{m+1})^2+y^2]/\sigma^2}
		\end{split}
		\right),
	\end{equation}
	where $p$ is the array depth, $\sigma$ is the waveguide width, $x_{m\pm1}=x_m \pm d \pm r\sin(\omega z)$ are the coordinates of the two outermost waveguides in each trimer that oscillate along $z$,  $x_m=3md$ are the coordinates of central waveguides in each trimer that in our case remain straight, $m$ is the index of the cell, $r$ is the modulation amplitude, and $\omega=2\pi/Z$ is the spatial oscillation frequency. One unit cell of such array has the width $3d$. Here we adopt the parameters $d=3.3$ ($33\, \mu \rm m$ spacing between neighbouring waveguides at $r=0$), $p=4.3$ (refractive index modulation depth $\delta n \sim 4.5\times 10^{-4}$), $\sigma=0.5$ ($5~\mu \rm m$-wide waveguides), and $Z=29$ (array period $\sim 33\, \rm mm$) typical for fs-laser written arrays at $\lambda=800\,\rm nm$~\cite{kirsch.np.17.995.2021, kartashov.prl.128.093901.2022, arkhipova.prl.130.083801.2023, tan.ap.3.024002.2021, li.ap.4.024002.2022, lin.us.2021.9783514.2021}. Further we consider sufficiently large finite arrays consisting of $11$ trimers.	
	
	\begin{figure}[htbp]
		\centering
		\includegraphics[width=1\columnwidth]{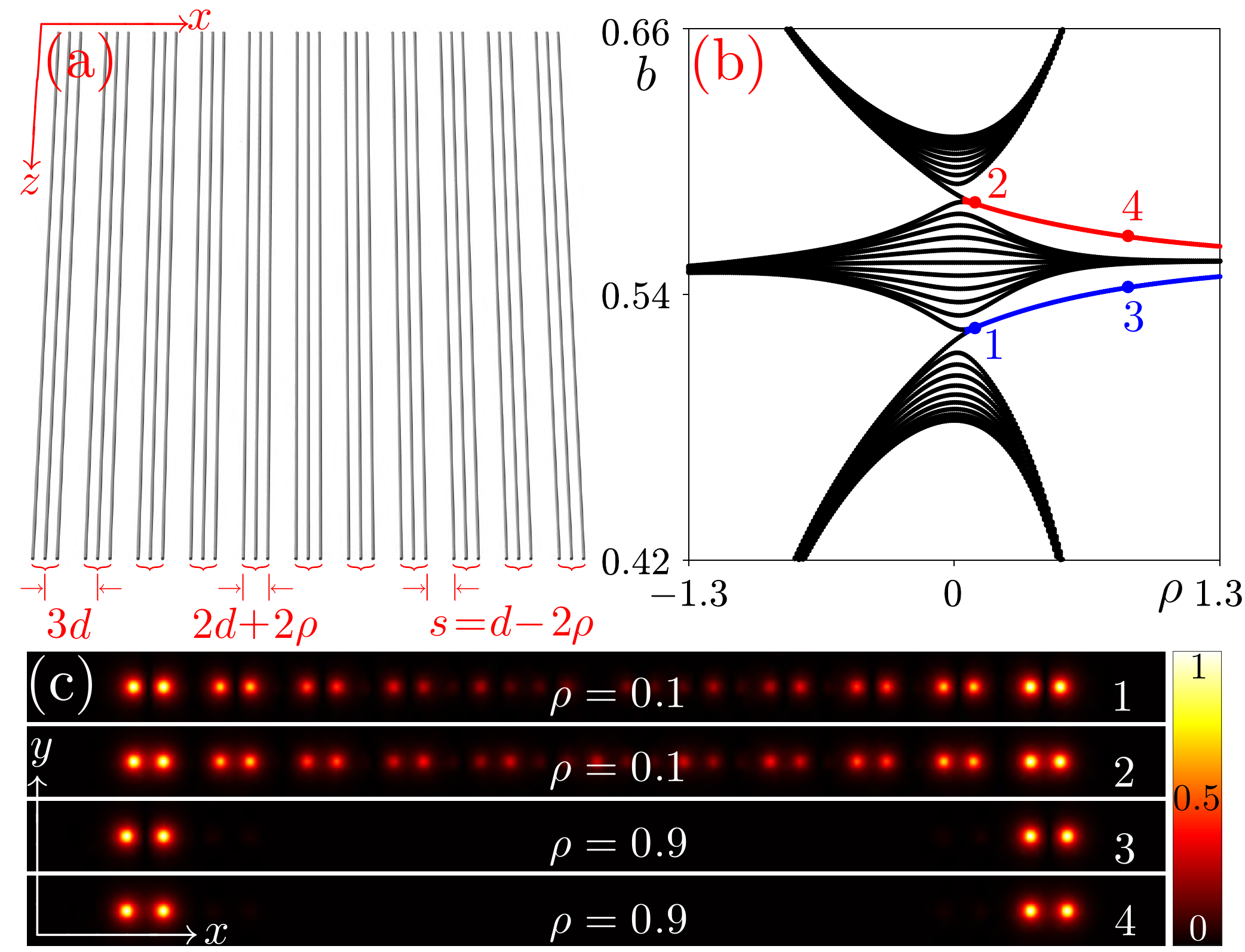}
		\caption{(a) Schematic illustration of the array consisting of trimers with straight waveguides. Each unit cell is indicated by a red bracket. The width of the unit cell is $3d$. The separation between central and outermost sites in each trimer is equal to $d+\rho$, where $\rho$ can be positive or negative. The separation between two closest sites of two neighbouring unit cells is $s=d-2\rho$. (b) Spectrum of the array with straight waveguides $b$ as a function of waveguide shift $\rho$. Red curves are $\pi$ modes and black curves are bulk states. (c) Amplitude profiles $|\psi|$ of $\pi$ eigenstates corresponding to the dots in (b). Eigenstates are shown in the window $-65\le x \le 65$ and $-4\le x \le 4$.}
		\label{fig1}
	\end{figure}
	
	\subsection{Spectrum of the array with straight waveguides}\label{sec_straight}
	
	To understand the mechanism of formation of topological edge states in Floquet array with oscillating waveguides it is instructive to consider first linear spectrum of static structure with straight waveguides [such structure is schematically depicted in Fig.~\ref{fig1}(a)]. To this end, we replace the oscillating term $r \textrm{sin}(\omega z)$ in the expression for coordinates $x_{m\pm1}$ of oscillating waveguides in Eq.~(\ref{eq2}) with constant shift $\rho$ and study transformation of the spectrum as a function of $\rho$. To obtain the spectrum, we substitute the ansatz $\psi(x,y,z)=\phi(x,y) e^{ibz}$, where $b$ is the propagation constant, into linear version of Eq.~(\ref{eq1}), and solve the resulting eigenvalue problem
	\begin{equation}\label{eq3}
		b\phi = \frac{1}{2} \left( \frac{\partial^2}{\partial x^2} + \frac{\partial^2}{\partial y^2} \right) \phi + \mathcal{R}(x,y) \phi,
	\end{equation}
	numerically using the plane-wave expansion method. The dependence of propagation constants of all modes of this static structure on $\rho$ is displayed in Fig.~\ref{fig1}(b). 
	This spectrum illustrates that two pairs of topological edge states appear in two gaps if $\rho>0$, as indicated by the red and blue curves, while at $\rho<0$ the structure is topologically trivial and no edge states emerge. Topological properties of such extended SSH-like arrays are described in previous works, see for instance~\cite{midya.pra.98.043838.2018, zhang.oe.29.42827.2021}. 
	The spectrum of static structure in Fig.~\ref{fig1}(b) implies that whether the array supports topological states or not is determined by waveguide shift $\rho$ which is fixed in this structure. For an oscillating arrays considered below this shift will change dynamically as $\rho=r\sin(\omega z)$ meaning that the array will actually periodically switch between ``instantaneous'' topological and nontopological phases depending on the value of $z$, spending in each phase exactly half of the longitudinal period $Z$.
	We show profiles of two edge states in the upper gap (labeled as 2 and 4) and another two edge states in the lower gap (labeled as 1 and 3) of static array in Fig.~\ref{fig1}(c). Since the states labeled 1 and 2 are closer to the bulk band, 
	they are more extended. Edge states in the upper and lower gaps differ in their phase structure: bright spots in the outermost two sites are in-phase in the state from upper gap, and they are out-of-phase in the state from lower gap. This picture agrees with previous findings~\cite{kartashov.prl.128.093901.2022}.
	
	\subsection{Spectrum of the Floquet trimer array}\label{sec_floquet}
	
\begin{figure}[htbp]
		\centering
		\includegraphics[width=1\columnwidth]{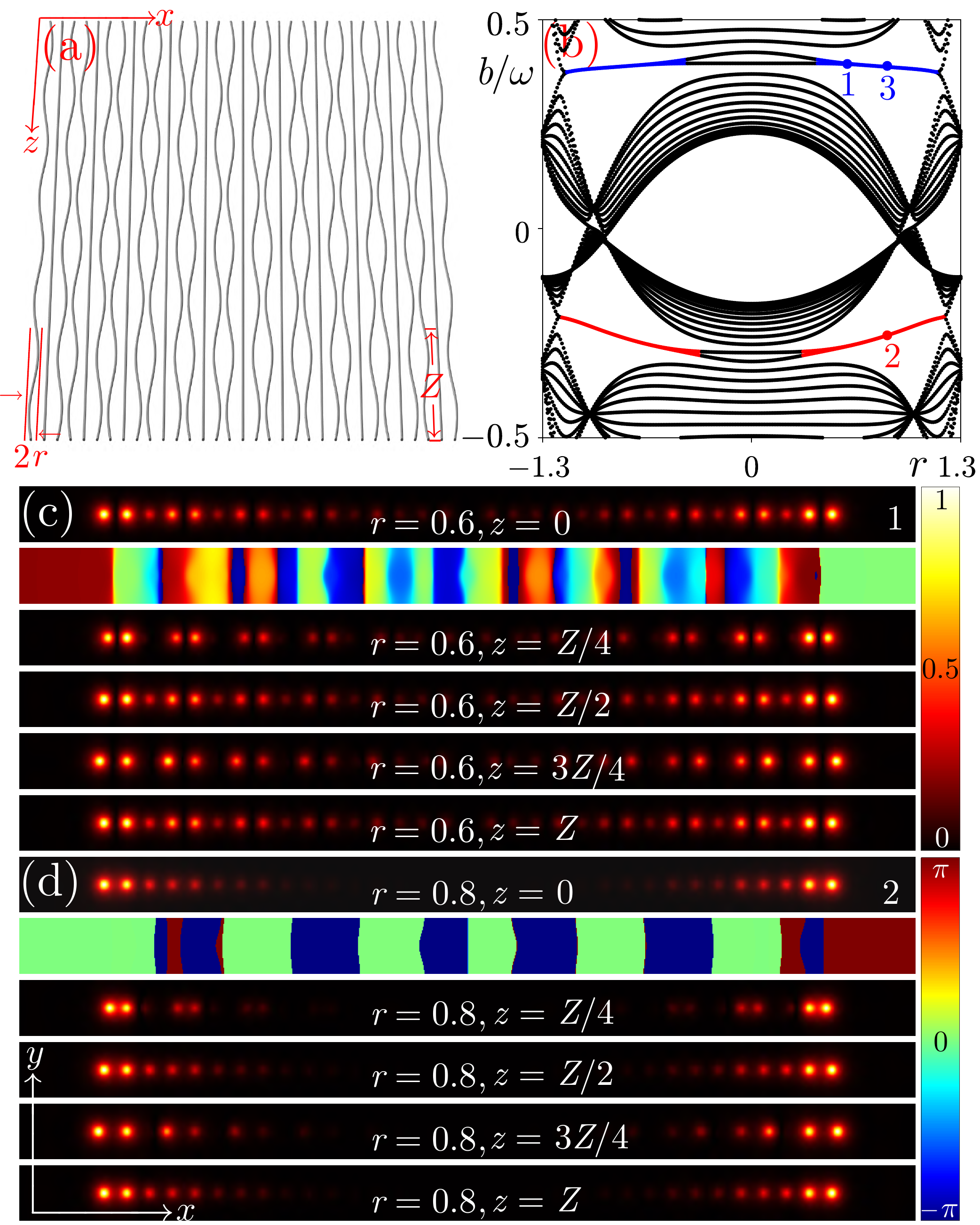}
		\caption{(a) Schematic illustration of the array of trimers with longitudinally oscillating (with amplitude $r$ and period $Z$) outermost waveguides in each unit cell. The central waveguide in each unit cell is unmodulated. The waveguides at the input facet $z=0$ are equally separated. (b) Quasi-propagation constant spectrum of the modulated trimer array $b$ as a function of amplitude $r$, where $\omega=2\pi/Z$ is the modulation frequency. Colored lines correspond to the Floquet edge states. (c) Field modulus distributions $|\psi|$ of the Floquet edge state at $r=0.6$ [dot 1 in (b)]at typical distances within one longitudinal period. (d) Field modulus distributions $|\psi|$ of the Floquet edge state at $r=0.8$ [dot 2 in (b)] at typical distances. Panels in (c,d) are shown in the window $-65\le x \le 65$ and $-4\le x \le 4$.
		In (c,d) the field modulus is shown with the red-black colormap, while the phase distribution for the field at $z=0$ is shown with the red-green-blue colormap.}
		\label{fig2}
	\end{figure}

	We now consider linear spectrum of modulated array. To this end we take the structure, where centers of two outermost waveguides in each trimer periodically oscillate with distance $z$, as shown in Fig.~\ref{fig2}(a), while central waveguide in each trimer remains straight. This is one of the simplest types of modulation that one can consider, for which the structure always remains $C_2$ symmetric, clearly distinct from previously considered types of modulations~\cite{wu.prr.3.023211.2021}. We select the phase of oscillations such that at $z=0$ all waveguides in the structure are equally separated, but this, of course, has no effect on the linear spectrum of the system. Notice that our structure is also clearly different from modulated SSH structure consisting of dimers considered in~\cite{song.lpr.15.2000584.2021}. Again, we substitute the ansatz $\psi(x,y,z)=\phi(x,y,z) e^{ibz}$, where now $\phi(x,y,z)=\phi(x,y,z+Z)$ is the $Z$-periodic function, and $b$ is the quasi-propagation constant, into linear version of Eq.~(\ref{eq1}) to obtain
	\begin{equation}\label{eq4}
		b\phi = \frac{1}{2} \left( \frac{\partial^2}{\partial x^2} + \frac{\partial^2}{\partial y^2} \right) \phi + \mathcal{R}(x,y,z) \phi + i \frac{\partial \phi}{\partial z},
	\end{equation}
	which is different from Eq.~(\ref{eq3}). The quasi-propagation constant spectrum of the finite Floquet array consisting of $11$ trimers can be obtained using ``propagation and projection'' method~\cite{leykam.prl.117.013902.2016,ivanov.apl.4.126101.2019}. In this method, at the first step we calculate the eigenvalues $b_j$ and eigenstates $\psi_j^{\rm in}$ of the optical potential ${\mathcal R}(x,y,z=0)$ with ${j\in[1,N]}$ and $N$ being the total number of waveguides. Next, we propagate all such ``static'' eigenstates $\psi_j^{\rm in}$ over one period $Z$ in modulated structure with refractive index distribution governed by ${\mathcal R}(x,y,z)$ to obtain output field distributions $\psi_j^{\rm out}$. We then calculate the matrix of projections with elements $U_{mn}=\langle \psi_m^{\rm in},\psi_n^{\rm out} \rangle$, whose eigenvalues are the Floquet exponents $e^{iZb_n}$ allowing to extract $b_n$ -- quasi-propagation constants of the $n$-th mode. To construct Floquet modes of the structure, for each $b_n$ we find the index $\ell_n$ of the maximum element of the corresponding eigenvector $V_n$ of $U_{mn}$. The Floquet eigenmode of ${\mathcal R}(x,y,z)$ with any index $n$ can then be constructed as $\psi_n = \sum_{j=1}^N \psi_j^{\rm in} V_j(\ell_n)$, and such modes satisfy the condition $\psi_n(z)=\psi_n(z+Z)$.
	
	In our case for $Z=29$ the width of the first longitudinal Brillouin zone (i.e, the modulation frequency $\omega$) $2\pi/Z \approx 0.22$ may be smaller than the width of the spectrum of static structure (considered for sufficiently large $\rho$ values), and on this reason the quasi-propagation constant spectrum may fold with increase of the amplitude $r$ of waveguide oscillations. This spectrum within the first longitudinal Brillouin zone is illustrated in Fig.~\ref{fig2}(b). The spectrum is periodic in $b$ with period $\omega$. One can see that longitudinal modulation leads to opening of \textit{two} topological gaps, where Floquet edge states indicated by the red and blue lines appear in proper range of modulation amplitudes $r$, even though the array undergoes periodic transitions between ``instantaneous'' topologically trivial and nontrivial configurations along the $z$ direction. Taking into account periodicity of the spectrum in $b$, one may conclude that upper/lower topological gaps with edge states in Floquet spectrum of Fig.~\ref{fig2}(b) qualitatively correspond to lower/upper topological gaps in static spectrum in Fig.~\ref{fig1}(b).
	
	At the same time, the most crucial difference between dynamic system considered here and previously addressed static trimer structures with straight waveguides (see Ref.~\cite{kartashov.prl.128.093901.2022}), is that here oscillations of two waveguides in each unit cell lead to periodic transformation of the array between instantaneously topological and nontopological phases, when it spends in each phase exactly \textit{half} of the longitudinal period $Z$. Under these conditions, the emergence of topological Floquet edge states in this system that remain localized near the edge for all distances z is controlled by the amplitude and period of longitudinal waveguide oscillations (rather than by constant for all $z$ shift of waveguides in static trimer array). Under these conditions simultaneous emergence of two topological gaps in Floquet spectrum with different types of edge states (and, consequently, of two different types of solitons that can bifurcate from such states) in them cannot be anticipated \textit{a priory} and is a surprising result.
	
	We choose one Floquet state from each band gap and display the corresponding field modulus distributions at typical propagation distances within one longitudinal period in Figs.~\ref{fig2}(c) and \ref{fig2}(d). Our array is symmetric at all distances $z$, and due to this in linear case edge states appear simultaneously at both left and right edges of the structure. Since the state numbered 1 with $r=0.6$ is closer to the bulk band, its localization is worse than that of the state numbered 2 with $r=0.8$. Even though the waveguide array is in trivial phase when $z\in (0,Z/2)$ and in topological phase when $z\in(Z/2,Z)$, the power of the Floquet state remains concentrated near the edges. From the profiles in Figs.~\ref{fig2}(c) and \ref{fig2}(d), it can be recognized that two outermost spots are approximately in-phase for the state numbered 2 and they are out-of-phase for the state 1. Such Floquet edge states exhibit considerable reshaping and amplitude oscillations in the course of propagation, exactly restoring their profiles after each period $Z$ (radiation is minimal for our parameters). Since waveguides in the array become equally spaced not only at $z=0$, but also at $z=Z/2$, the intensity distributions at these distances are similar, but phase structures are different.
	
	It should be also pointed out that inside each gap Floquet edge states actually emerge in pairs (see Fig. \ref{fig2}(b)) that become practically degenerate with increase of the amplitude of waveguide oscillations $r$. These pairs correspond to in-phase states near opposite edges, or to out-of-phase states near opposite edges (do not mix this with different phase relations between neighboring spots in states from different gaps discussed above). Because states in a given gap become practically degenerate already at $r\sim 0.5$ their linear combinations give Floquet state concentrated near one edge only. Consequently, in the presence of nonlinearity one can get solitons concentrated also near only one edge (see below).
	
	In Fig.~\ref{fig2}(a), the Floquet trimer array is $C_2$ symmetric, since two outer waveguides in each unit cell oscillate out-of-phase. At the same time, one can also realize the structure where these two waveguides oscillate in-phase, in which case the array lacks the $C_2$ symmetry. However, in this case simulations show that no edge states appear in Floquet spectrum of the system, i.e. it becomes topologically trivial. Thus, while there are many possible ways to design the Floquet trimer lattice, the design used in Fig.~\ref{fig2} seems to be among the simplest ones allowing to realized nontrivial Floquet topology.
	
	\subsection{Topological origin of the Floquet edge states}
	
	Topological characterization of this system can be performed by calculating the Zak phase for the bands of periodic (non-truncated) structure. If we denote the $z$-dependent intracell coupling strength (i.e. coupling strength between the outermost and central waveguides of the trimer) by $w_z$ and the intercell coupling strength (between left and right outermost waveguides in neighboring trimers) by $v_z$, in the frames of the tight-binding approximation accounting for coupling between neighbouring waveguides only, one can introduce the following Floquet evolution operator $U(k,Z)$ that can be formally expressed via effective Hamiltonian $H_{\rm eff}$ of the Floquet trimer system~\cite{kitagawa.prb.82.235114.2010, rudner.prx.3.031005.2013, fruchart.prb.93.115429.2016}
	\begin{equation}
		U(k,Z)=\mathcal{Z}
		e^{-i\int_0^Z H(k,z) dz} = e^{-iH_{\rm eff}(k) Z},
	\end{equation}
	where $\mathcal{Z}$ is the time-ordering operator, and $H(k,z)$ is the ``instantaneous'' Hamiltonian of the periodic system
	
	\begin{equation}
		H(k,z)=
		\begin{bmatrix}
			0 & w_z & v_z e^{-i3dk} \\
			w_z & 0 & w_z \\
			v_z e^{i3dk} & w_z & 0 
		\end{bmatrix}.
	\end{equation}

	Here, $w_z$ and $v_z$ periodically change with propagation distance, and $k\in[0,K]$ is the Bloch momentum with $K=2\pi/3d$ being the width of the first transverse Brillouin zone. If we use $|u(k)\rangle$ to represent the eigenvector of the effective Hamiltonian $H_{\rm eff}$, the Zak phase~\cite{zak.prl.62.2747.1989} of the bulk band can be calculated as
	\begin{equation}\label{eq7}
		\gamma=i \oint \langle u(k) | \frac{d}{dk} | u(k) \rangle dk.
	\end{equation}
	One finds that for our type of modulation, the Zak phase of the middle band of the effective Hamiltonian [that corresponds in continuous system to the band located close to the bottom of the first longitudinal Brillouin zone just below red lines in Fig.~\ref{fig2}(b)] is $2\pi$, while for two other bands [corresponding to two bands above red lines and below blue lines in Fig.~\ref{fig2}(b) clearly resolvable in spectrum at oscillation amplitudes $r<1$] the Zak phase is $\pi$, that is consistent with arrangement of edge states in Fig.~\ref{fig2}(b) appearing between bands with different topological indices.
	
	\section{Floquet edge solitons}\label{nonlinear}
	
	As demonstrated in Sec.~\ref{sec_floquet}, our system supports two types of linear Floquet edge states with different phase structure in two different topological gaps. In this section, we investigate Floquet edge solitons that can bifurcate from these linear edge states. Now, we use the ansatz $\psi(x,y,z)=\phi(x,y,z) e^{ibz}$ in the Schr\"odinger equation (\ref{eq1}) and keep nonlinear term, that yields the equation
	\begin{equation}\label{eq5}
		b\phi = \frac{1}{2} \left( \frac{\partial^2}{\partial x^2} + \frac{\partial^2}{\partial y^2} \right) \phi + \mathcal{R}(x,y,z) \phi + i \frac{\partial \phi}{\partial z} + |\phi|^{2} \phi,
	\end{equation}
	where $b$ is the quasi-propagation constant of the Floquet edge soliton. Such states can be obtained using iterative method that involves the following steps~\cite{lumer.prl.111.243905.2013,ren.csf.166.113010.2023}:
	(1) We propagate the linear Floquet edge state $\psi_{\rm fl}^{\rm in}$ with a given power $P$ according to Eq.~(\ref{eq1}) to obtain the dynamical lattice modified by the nonlinearity, i.e. $\mathcal{R}_{\rm fl}=\mathcal{R}+|\psi_{\rm fl}^{\rm in}|^2$. 
	(2) We propagate all linear eigenstates $\psi^{\rm in}_j$ of $\mathcal{R}$ that include $\psi_{\rm fl}^{\rm in}$ in the modified dynamical lattice $\mathcal{R}_{\rm fl}$ for a whole period $Z$ and obtain corresponding output distributions $\psi^{\rm out}_j$. 
	(3) We calculate the projection $U_{mn} = \langle \psi^{\rm in}_m, \psi^{\rm out}_n \rangle$, whose eigenvalues are Floquet exponents $e^{iZb_n}$. 
	(4) For each $b_n$, we find the index $\ell_n$ of the maximum element of the corresponding eigenvector $V_n$. Eigenstates $\psi_n^{\rm re}$ of $\mathcal{R}_{\rm fl}$ can then be constructed as $\psi_n^{\rm re}=\sum_{q=1}^N \psi_q^{\rm in} V_q(\ell_n)$. 
	(5) We pick out the modified edge state $\psi_{\rm fl}^{\rm re}$ from $\psi_n^{\rm re}$ and normalize it in accordance with selected power $P$. 
	(6) The steps (1)-(5) are repeated until the difference between $\psi_{\rm fl}^{\rm re}$ and $\psi_{\rm fl}^{\rm in}$ reduces below required small level.
	
	\begin{figure}[htbp]
		\centering
		\includegraphics[width=\columnwidth]{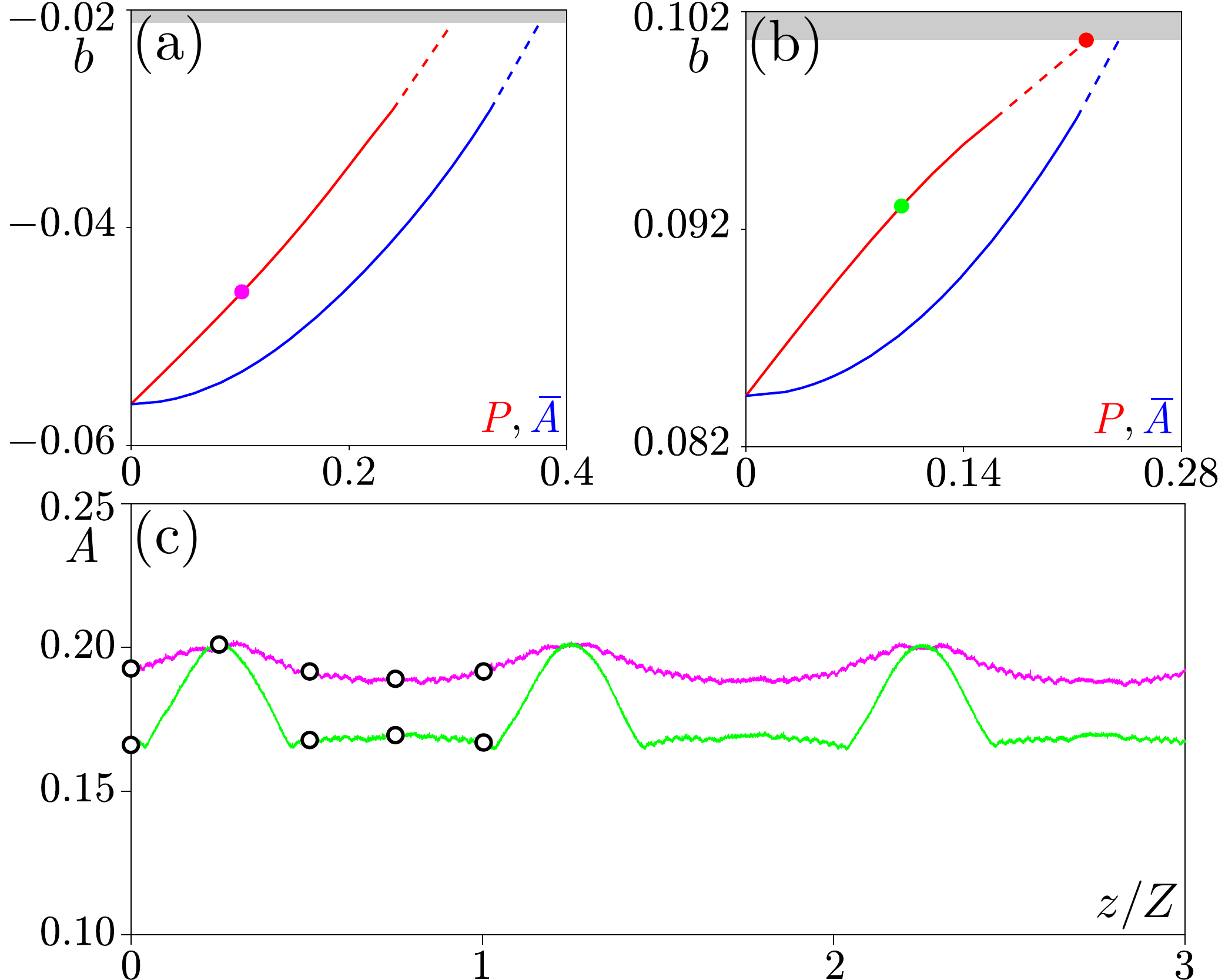}
		\caption{(a) Dependence of the quasi-propagation constant $b$ and averaged peak amplitude $\bar A$ on the power $P$ for the Floquet soliton family at $r=0.8$ bifurcating from linear mode corresponding to the dot 2 in Fig.~\ref{fig2}(b). The shaded region represents the bulk band; solid and dashed lines represent stable and unstable solitons, respectively. 
			(b) Similar dependencies for soliton family bifurcating from linear Floquet state numbered 3 in Fig.~\ref{fig2}(b), also at $r=0.8$. 
			(c) Peak amplitudes of the Floquet solitons during propagation. The magenta and green curves correspond to the magenta dot ($P=0.10$) and green dot ($P=0.10$) in panels (a) and (b), respectively.}
		\label{fig3}
	\end{figure}
	
	Quasi-propagation constant $b$ of such solitons and their averaged peak amplitude $\bar A$  are therefore controlled by their power $P$:
	\begin{equation}
		P=\iint |\psi|^2 dx dy,\quad \bar A=\frac{1}{Z}\int_z^{z+Z} A(z) dz,
	\end{equation}
	where $A(z)=\max\{ |\psi| \}$ is the peak amplitude of the Floquet edge soliton. In Figs.~\ref{fig3}(a) and \ref{fig3}(b), we show the dependencies $b(P)$ and $b(\bar A)$ for Floquet solitons bifurcating from linear Floquet edge states numbered 2 and 3 in Fig.~\ref{fig2}(b) at $r=0.8$. One finds that for both types of solitons the quasi-propagation constant increases with the power and gradually approaches the bottom edge of the bulk band [indicated by gray shaded regions in Figs.~\ref{fig3}(a) and \ref{fig3}(b)], where we stop our calculations, because in the band such states couple with bulk modes. Floquet edge solitons are stable when they correspond to parts of families shown by solid curves in Figs.~\ref{fig3}(a) and \ref{fig3}(b), and they become unstable when they approach the bulk band, as indicated by dashed curves.
		
	\begin{figure*}[htbp]
		\centering
		\includegraphics[width=\textwidth]{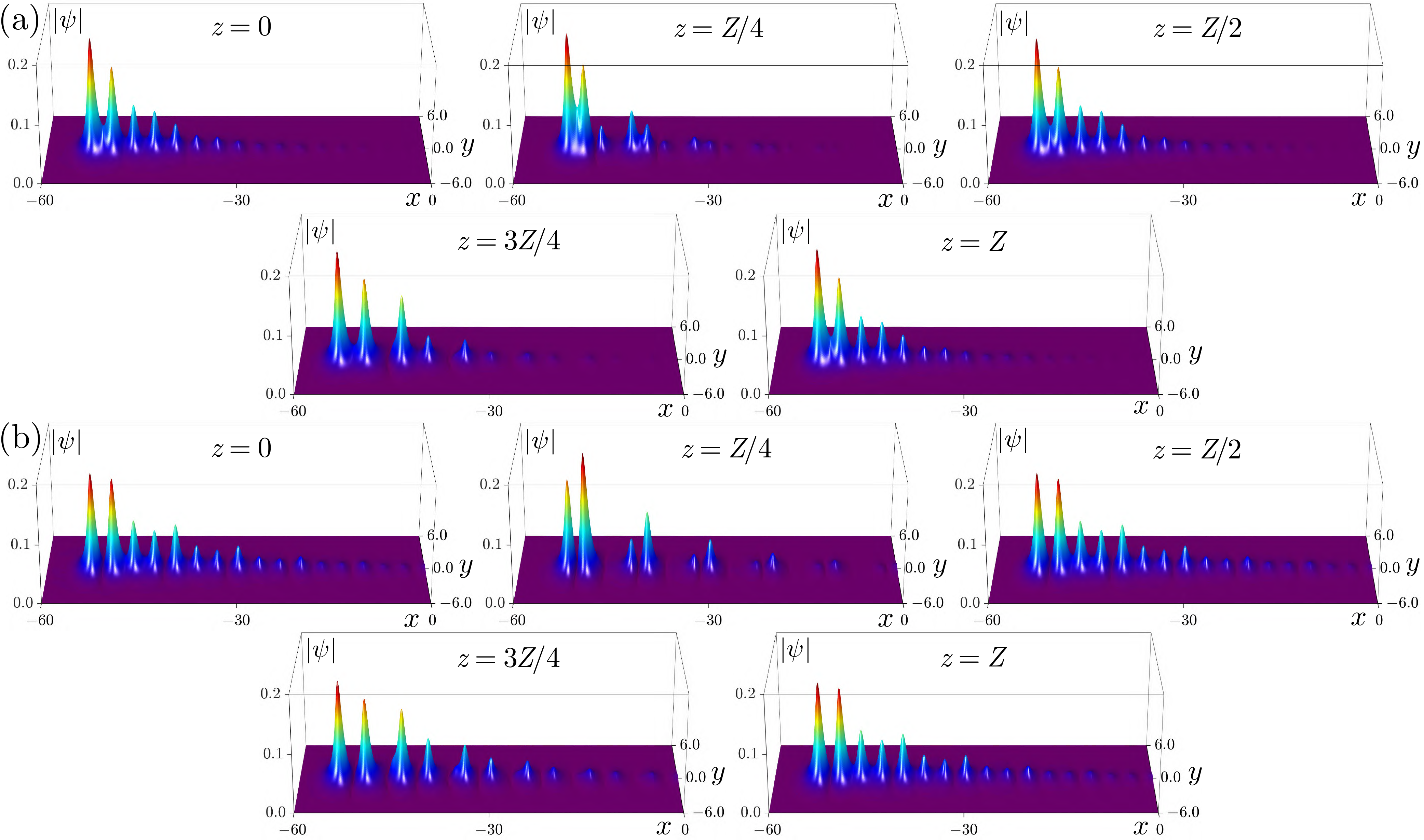}
		\caption{(a) Field modulus distributions $|\psi|$ of the nonlinear Floquet edge state at different distances from the magenta curve in Fig.~\ref{fig3}(c). 
			(b) Field modulus distributions $|\psi|$ of the nonlinear Floquet edge state at different distances from the green curve in Fig.~\ref{fig3}(c).}
		\label{fig4}
	\end{figure*}
	
	\begin{figure}[htbp]
		\centering
		\includegraphics[width=1\columnwidth]{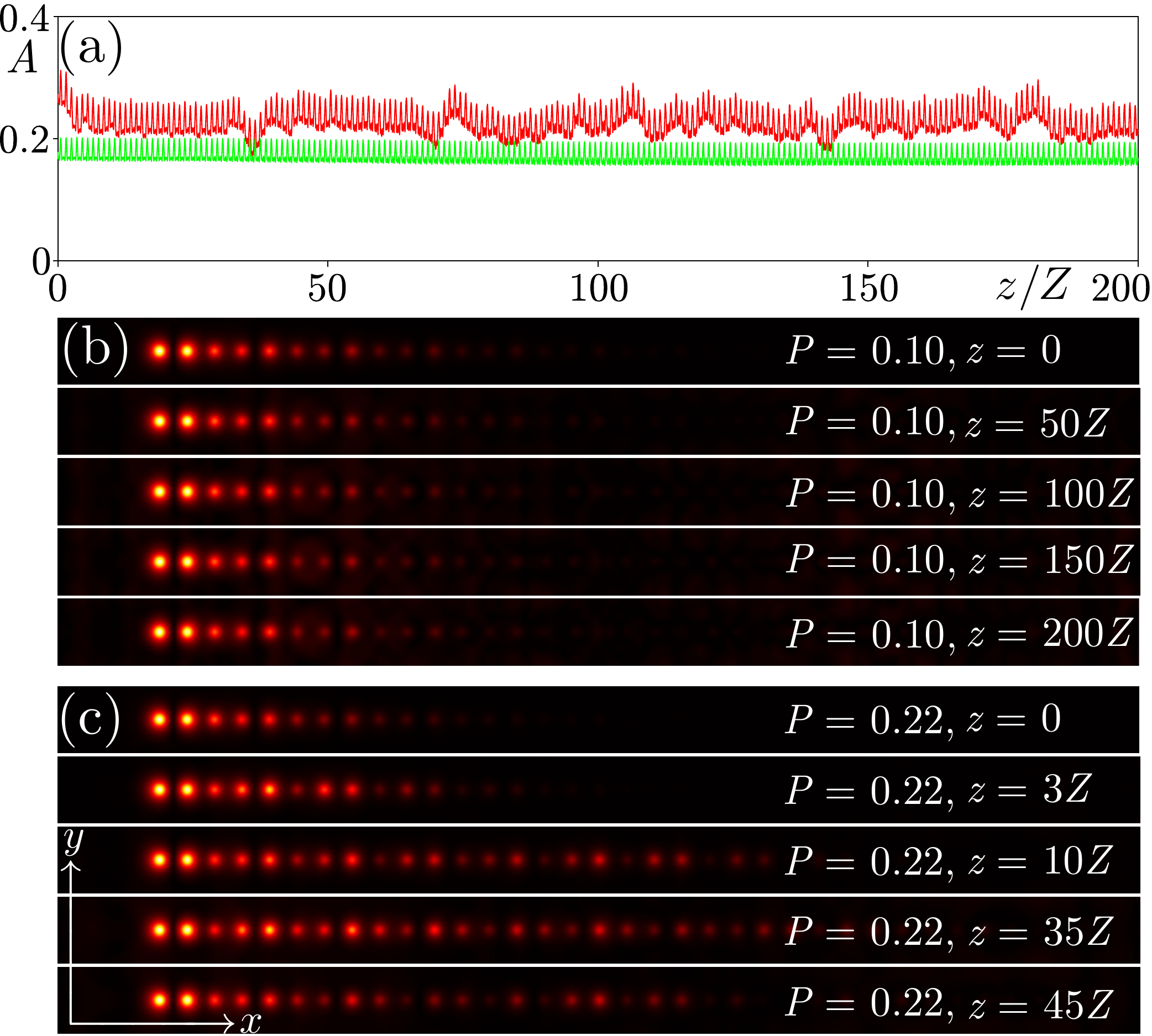}
		\caption{(a) Peak amplitude of the perturbed Floquet edge soliton vs propagation distance. The green and red curves correspond to the green dot ($P=0.10$) and the red dot ($P=0.22$) in Fig.~\ref{fig3}(b), respectively. (b) Field modulus distributions in stable Floquet edge soliton from the green curve in (a) at different distances. (c) Unstable Floquet edge soliton from the red curve in (a) at different distances. Panels in (b,c) are shown in the window $-65\le x \le 65$ and $-4\le x \le 4$.}
		\label{fig5}
	\end{figure}
	
	Floquet edge solitons in modulated trimer arrays are nonlinear dynamically oscillating states that periodically reproduce their shapes with period $Z$. To illustrate this, we choose two solitons with $P=0.1$ from both gaps corresponding to the magenta dot in Fig.~\ref{fig3}(a) and to the green dot in Fig.~\ref{fig3}(b), and show in Fig.~\ref{fig3}(c) with lines of similar color the evolution of their peak amplitudes with distance on three longitudinal periods of the array. Clearly, the peak amplitude $A$ exhibits periodic behavior. It is important to stress, that such stable states do not decay even at distances of the order of hundreds of longitudinal periods (see an example below).
	
	In Figs.~\ref{fig4}(a) and \ref{fig4}(b) we depict representative field modulus distributions in solitons at five typical distances $z=0$, $z=Z/4$, $z=Z/2$, $z=3Z/4$ and $z=Z$ corresponding to the open dots in Fig.~\ref{fig3}(c). The exact replication of soliton profile at $z=0$ and $z=Z$ is obvious, the $|\psi|$ distribution at $z=Z/2$ is also very similar. Since the separation between waveguides is the smallest at $z=Z/4$, the overlap of the left two spots of the nonlinear Floquet edge state is the most pronounced at this distance. At the same time, at $z=3Z/4$ the separation between spots is maximal. Comparing field modulus distributions in two solitons presented in Figs.~\ref{fig4}(a) and \ref{fig4}(b) one can notice that such solitons clearly inherit phase structure from linear edge states: in Fig.~\ref{fig4}(a) the two left spots are approximately in-phase that leads to constructive interference of the fields between two waveguides, while in Fig.~\ref{fig4}(b) the spots are out-of-phase and intensity vanishes between two outermost waveguides.
	
	In addition, we also inspect the propagation dynamics of perturbed stable and unstable Floquet edge solitons over large distances. We consider states corresponding to the green dot ($P=0.1$) and red dot ($P=0.22$) in Fig.~\ref{fig3}(b), superimpose a small random noise $\delta(x,y)\in[-0.05,0.05]$ on the exact Floquet state at $z=0$, so that initial field distribution is given by $\phi(x,y,0) [1+\delta(x,y)]$, and propagate it in the frames of Eq.~(\ref{eq1}) up to several hundreds of longitudinal periods. The corresponding dependence of peak amplitude on distance is displayed in Fig.~\ref{fig5}(a). Stable state (green curve) shows exactly periodic evolution, Fig.~\ref{fig5}(b) shows corresponding field modulus distributions confirming that soliton shape taken at $z=nZ$, where $n$ is an integer, does not change and no radiation is present. In contrast, the amplitude of the unstable state [red curve in Fig.~\ref{fig5}(a)] is periodic only over first couple of periods, but then it undergoes irregular oscillations. This state exhibits considerable radiation into the bulk and eventually drastically broadens in the course of propagation, as shown in Fig.~\ref{fig5}(c).
	
\begin{figure}[htbp]
	\centering
	\includegraphics[width=1\columnwidth]{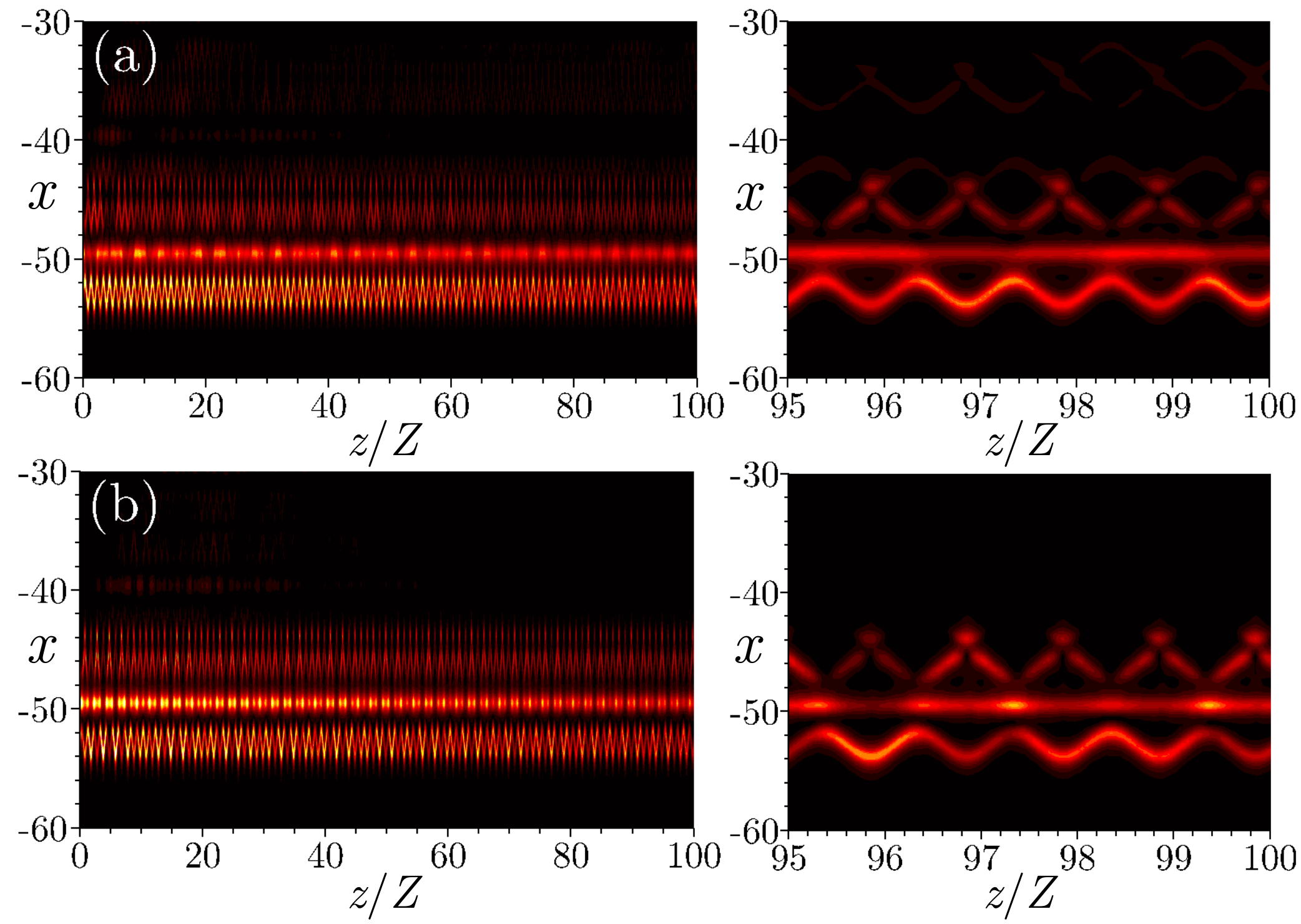}
	\caption{Dynamics of excitation of Floquet edge solitons by two in-phase (a) and out-of-phase (b) Gaussian beams (total power $P=0.2$) that are launched into two outermost waveguides of the Floquet trimer array with $r=1$.
		The right two panels are magnified propagation dynamics in $95Z\le z \le 100Z$.}
	\label{fig6}
\end{figure}
	
	Nonlinear Floquet states reported here can be excited dynamically by using proper input beams. To demonstrate this, we launched two Gaussian beams with $0$ or $\pi$ phase difference between them into two left waveguides of the Floquet trimer array and examine their long-distance propagation. Using two beams with proper phase difference maximizes overlap of the input field with one of the Floquet solitons from different gaps, either in-phase or out-of-phase one. Corresponding propagation dynamics is illustrated in Fig.~\ref{fig6}. As one can see, even though shedding of some radiation into the bulk is present because the inputs do not exactly match the profiles of solitons, two nonlinear states with clearly different dynamics and internal structure are excited, the light remains confined near the edge even at distances up to $100Z$, and the amplitude shows practically periodic oscillations indicating on the formation of the in-phase and out-of-phase Floquet edge solitons.

	\section{Conclusions}
	
	Summarizing, we have theoretically introduced co-existing Floquet edge states with different phase structures in two different topological band gaps in the spectrum of the one-dimensional Floquet trimer array. The appearance of these states is consistent with different topological indices (Zak phases) of three bulk bands in Floquet spectrum of corresponding periodic structure. We also predicted that Floquet edge solitons may bifurcate from both types of linear edge states in the presence of nonlinearity. These nonlinear Floquet edge solitons exhibit periodic behavior during propagation and can be dynamically stable. Our results illustrate rich opportunities for the realization of topologically nontrivial phases offered by periodic longitudinal modulations of shallow waveguiding structures, that may be potentially used in the design of Floquet topological lasers, higher-order Floquet topological insulators, and other new classes of topological devices.
	
	\section*{Acknowledgement}
	This research was funded by the National Natural Science Foundation of China (Grant Nos.: 12074308, U1537210),
	the research project FFUU-2021-0003 of the Institute of Spectroscopy of the Russian Academy of Sciences, and
	the Russian Science Foundation grant 21-12-00096.
	

%

	
\end{document}